\documentclass[prl,twocolumn,floatfix]{revtex4}
\usepackage{graphicx}
\usepackage{color}

\newcommand{\etal}{{\em et al.}}
\begin{document}

\title {Comparison of Source Images for protons, $\pi^-$'s and $\Lambda$'s in 6 AGeV Au+Au collisions}

\author{
P.~Chung$^{(1)}$, N.~N.~Ajitanand$^{(1)}$,
J.~M.~Alexander$^{(1)}$,
M.~Anderson$^{(6)}$, D.~Best$^{(7)}$,
F.P.~Brady$^{(6)}$, T.~Case$^{(7)}$, W.~Caskey$^{(6)}$,
D.~Cebra$^{(6)}$,
J.L.~Chance$^{(6)}$, B.~Cole$^{(12)}$, K.~Crowe$^{(7)}$,
A.~C.~Das$^{(3)}$, J.E.~Draper$^{(6)}$, M.L.~Gilkes$^{(1)}$,
S.~Gushue$^{(1,10)}$, M.~Heffner$^{(2,6)}$,
A.S.~Hirsch$^{(7)}$, E.L.~Hjort$^{(7)}$, L.~Huo$^{(14)}$,
M.~Justice$^{(5)}$,
M.~Kaplan$^{(9)}$, D.~Keane$^{(5)}$, J.C.~Kintner$^{(13)}$,
J.~Klay$^{(6)}$,
D.~Krofcheck$^{(11)}$,R.~A. ~Lacey$^{(1)}$,
J.~Lauret$^{(1)}$, M.A.~Lisa$^{(3)}$, H.~Liu$^{(5)}$,
Y.M.~Liu$^{(14)}$,
R.~McGrath$^{(1)}$, Z.~Milosevich$^{(9)}$, G.~Odyniec$^{(7)}$,
D.L.~Olson$^{(7)}$, S.~Panitkin$^{(5)}$,
N.T.~Porile$^{(8)}$, G.~Rai$^{(7)}$, H.G.~Ritter$^{(7)}$,
J.L.~Romero$^{(6)}$, R.~Scharenberg$^{(8)}$, B.~Srivastava$^{(8)}$, 
N.T.B~Stone$^{(7)}$, T.J.M.~Symons$^{(7)}$,
J.~Whitfield$^{(9)}$, R.~Witt$^{(5)}$, L.~Wood$^{(6)}$,
and W.N.~Zhang$^{(14)}$
                        \\  (E895 Collaboration) \\
and\\
D.~Brown$^{(2)}$, S.~Pratt$^{(4)}$,F.~Wang$^{(7)}$,
P.~Danielewicz$^{(4)}$ \\
}
\address{
$^{(1)}$Depts. of Chemistry and Physics,
SUNY \@ Stony Brook, New York 11794-3400 \\
$^{(2)}$Livermore National Laboratory, Livermore, CA 94550\\
$^{(3)}$Ohio State University, Columbus, Ohio 43210\\
$^{(4)}$National Superconducting Cyclotron Laboratory,\\
Michigan State University, East Lansing, MI  48824\\
$^{(5)}$Kent State University, Kent, Ohio 44242 \\
$^{(6)}$University of California, Davis, California, 95616 \\
$^{(7)}$Lawrence Berkeley National Laboratory,
Berkeley, California, 94720 \\
$^{(8)}$Purdue University, West Lafayette, Indiana, 47907-1396 \\
$^{(9)}$Carnegie Mellon University, Pittsburgh, Pennsylvania 15213\\
$^{(10)}$Brookhaven National Laboratory, Upton, New York 11973 \\
$^{(11)}$University of Auckland, Auckland, New Zealand \\
$^{(12)}$Columbia University, New York, New York 10027 \\
$^{(13)}$St. Mary's College, Moraga, California  94575 \\
$^{(14)}$Harbin Institute of Technology, Harbin, 150001 P.~R. China\\
 }
\date{\today}

\begin{abstract}
Source images are extracted from two-particle correlations constructed
from strange and non-strange hadrons produced in 6 AGeV Au + Au collisions.
Very different source images result from pp vs p$\Lambda$ vs $\pi^-\pi^-$
correlations. These observations suggest important differences in the
space-time emission histories for protons, pions and neutral strange baryons
produced in the same events.
\end{abstract}


\maketitle


Relativistic heavy ion collisions of 1-10~AGeV produce a
fireball of nuclear matter with extremely high baryon and energy
density\cite{stocker86}. The dynamical evolution of this fireball is
driven by such fundamental properties as the nuclear Equation of State (EOS)
and possibly by a phase transition, e.g., to a Quark Gluon Plasma (QGP)
\cite{rischke96,sor97,dan98,qm96,pin99}.
Two-particle correlation studies,
for various particle species, provide an important probe of the
space-time extent of this fireball\cite{lis00,E866,pratt93,Na44}.
Recent model calculations suggest that the time scale for freeze-out
of strange and multi-strange particles may be much shorter than that for 
non-strange particles\cite{hecke98,senger98},
implying a much smaller space-time emission zone for strange particles.
If this is indeed the case,
then correlation studies involving strange particles may serve as
important ``signposts" for dynamical back-tracking into the very early
stage of the collision where large energy densities are achieved \cite{wan99}.

In this letter we compare the source properties for protons, $\pi^-$'s and
$\Lambda$ hyperons extracted from $pp$, $\pi^-\pi^-$ and $p\Lambda$ correlation 
functions, as produced in 6 AGeV Au+Au collisions.
These data are unique in that they constitute the first measurement of $p\Lambda$ 
correlations. If $\Lambda$ hyperons are in fact emitted from a source with a 
smaller space-time extent, then between this and $\pi^-$ source broadening 
from resonance decays, one might naively expect an ordering of two-particle 
source sizes: $R_{p\Lambda} < R_{pp} < R_{\pi\pi}$.
On the other hand, at these energies the 3-dimensional $\pi^-$ radii exhibit $m_T$
scaling \cite{lis00,ahle02} and this should manifest itself in the angle-averaged 
$\pi^-$ sources.  Furthermore, since $m_T$ scaling can be ascribed to 
position-momentum correlations in the particle emission function 
\cite{mTscaling}, one might expect similar effects in the 
$pp$ and $p\Lambda$ sources. As we will show, neither an interpretation based solely
on position-momentum correlations nor on naive geometrical arguments can
fully account for the relative size of the three sources.

Traditional correlation analyses rely on the weakness of
Final-State Interactions (FSI).  With this, one may correct for FSI leaving a
correlation that is the Fourier Transform of the two-particle source function.
This approximation is mostly valid for pions; for massive and/or strongly
interacting particles, such as protons or $\Lambda$ hyperons, this
approximation breaks down.
Recently, Brown and Danielewicz have presented an 
imaging technique for analyzing two-particle correlations \cite{bro01}.
The technique actually {\em uses} the FSI, encoded in the
form of the particles' final state wavefunction, to extract the two-particle
source function {\em directly} \cite{bro01}.
The imaging technique has been used to address only a few data sets; Panitkin
\etal \cite{pan01} have shown that this approach gives source radii similar to
the conventional HBT approach (under the assumption of a Gaussian source)
for $\pi^-\pi^-$ pairs emitted in central collisions for 2-8 AGeV
Au+Au.  By contrast, Verde \etal\cite{verde01} have shown very different
results for $pp$ pairs from 75 AMeV $^{14}$N+$^{197}$Au.
For our purpose, the imaging technique's key feature is that one can
easily compare source functions across different species and
assess the different space-time scales relevant for each particle pair -- a
feature that has been mentioned \cite{bro01}, but never seriously utilized.

We use the imaging technique on $pp$, $\pi^-\pi^-$, and $p\Lambda$ pairs
to extract and compare the source properties for protons, $\pi^-$'s and
$\Lambda$ hyperons.
The measurements have been performed with the E895 detector at
the Brookhaven Alternating Gradient Synchrotron. Here, we
concentrate on the construction of the $p\Lambda$ and $pp$ correlations 
and the results of the imaging analysis. Details on the detector and its 
setup have been reported elsewhere \cite{pin99,chu00b,rai90}.

\begin{figure}
\includegraphics[width=0.45\textwidth]{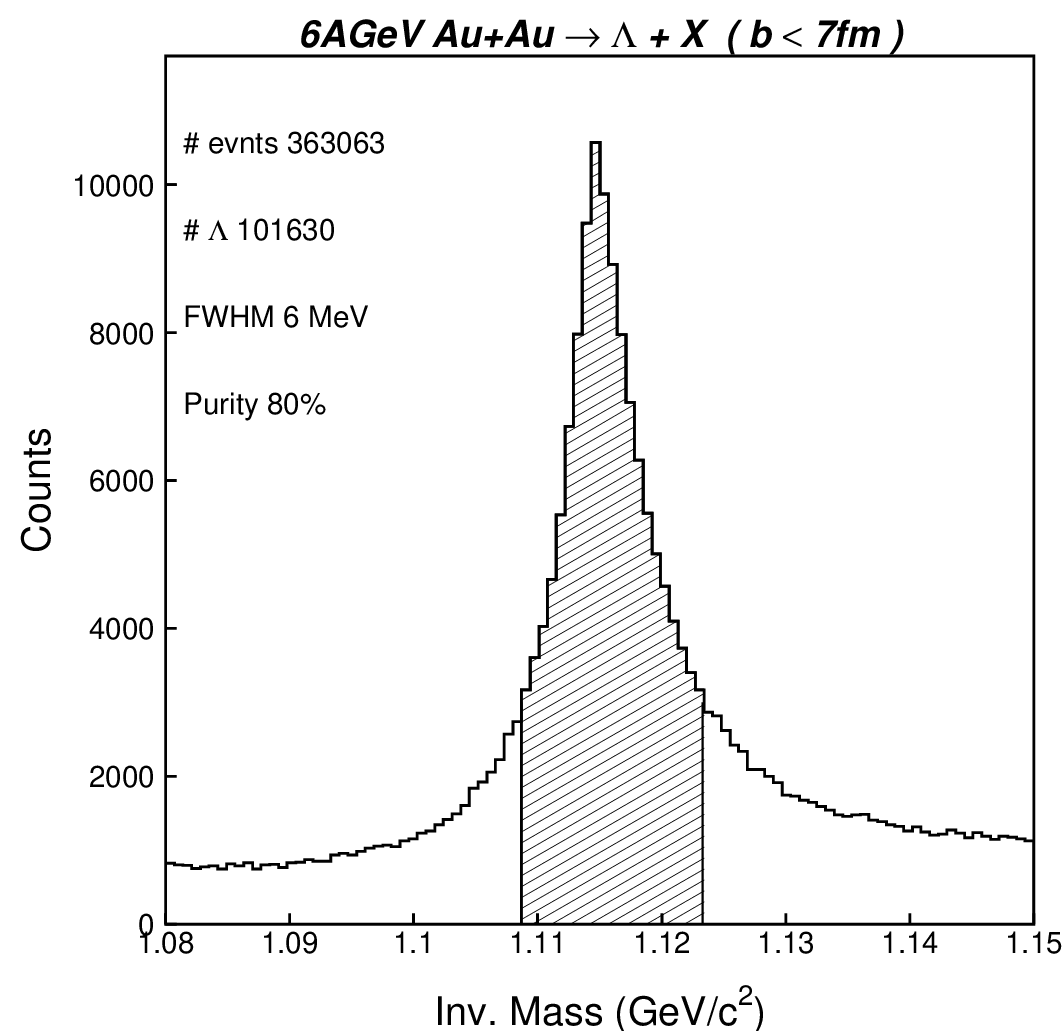}
\caption{$\Lambda$ invariant mass spectrum from semi-central
6AGeV Au+Au collisions. The hatched area depicts the $\Lambda$ mass gate
used to select the $\Lambda$'s for $\Lambda$p correlation analysis.
}
\label{lam_mass_cor}
\end{figure}

We reconstructed the $\Lambda$'s from the daughters of their charged
particle decay, $\Lambda \longrightarrow p + \pi^-$
(Branching Ratio $\simeq$ 64 \%), following the procedure outlined in
Refs.\cite{chu00b}.  Figure~\ref{lam_mass_cor}
shows the invariant mass spectrum for
$\Lambda$'s obtained in semi-central (upper 23\% of total inelastic
cross-section, b $\lesssim $7 fm)
6 AGeV Au + Au collisions. For the p$\Lambda$ correlation analysis, an
enriched sample ($\sim 80$\%) of $\Lambda$'s with
1.11 $\le M_\Lambda \le$ 1.122 GeV was used.

\begin{figure}
\includegraphics[width=0.45\textwidth]{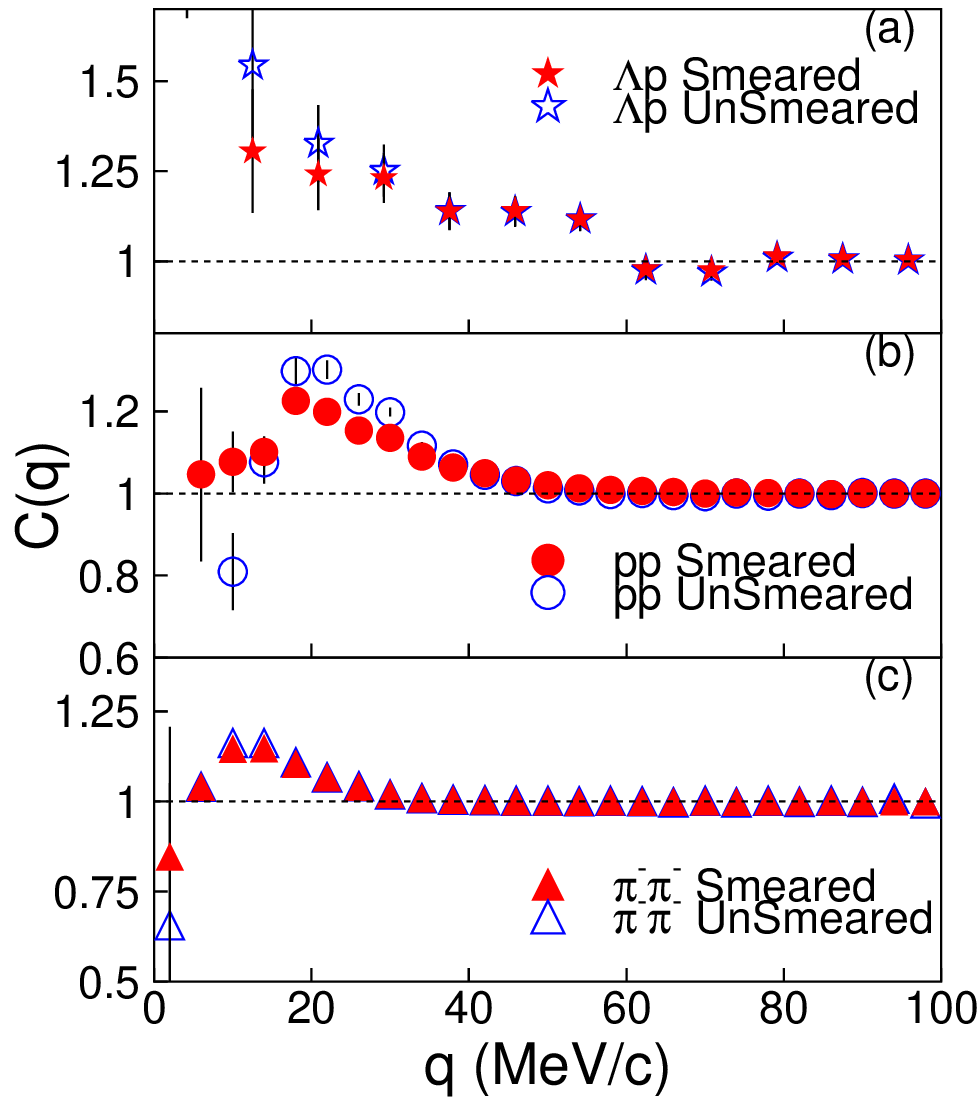}
\caption{Raw smeared (filled symbols) and corrected
unsmeared (open) correlation functions, C(q), for (a) $\Lambda$p,
(b) pp and (c) $\pi^-\pi^-$ from 6AGeV Au+Au collisions (b$<$7fm).
}
\label{smr_unsmr}
\end{figure}

Figures~\ref{smr_unsmr}(a),(b),(c) show the correlation
functions, $C(q)$, obtained by taking the ratio of foreground to background
distributions in relative-momentum for $\Lambda$p, pp and $\pi^-\pi^-$ pairs
respectively. Here, $q=\frac{1}{2}\sqrt{-(p_1-p_2)^2}$ is half of the
relative momentum between the two particles in the pair c.m. frame.  We applied
no explicit gates on transverse momentum
and rapidity except those implicit in the tracking
acceptance. The mean transverse momenta $\left<P_t\right>$ for particles from
pairs with $q<50$MeV  were 0.3GeV and 0.12GeV for pp and $\pi^-\pi^-$ pairs
and the mean rapidities $\left<y\right>$ were -0.33 and 0.1 .
$\Lambda$p pairs with $q<50$MeV have $\left<P_t\right>$ = 0.46GeV and
$\left<y\right>$=-0.18 for $\Lambda$'s
and $\left<P_t\right>$ = 0.38GeV and $\left<y\right>$=-0.18 for protons.

We constructed the numerator (or foreground) distribution
from pairs of particles from the same event, and
obtained the denominator (or background) distribution by pairing
particles from different events. We used a track-merging filter similar to
that outlined in Ref.\cite{lis00} to eliminate possible distortions
resulting from track-merging effects
in the TPC. For each correlation function, we used the accepted range of
particle multiplicities to specify impact parameters
b $<$ 7 fm. This range was chosen to optimize the
statistical significance for p$\Lambda$ pairs.

We utilized approximately 100,000 $\Lambda$'s (80\% purity) to yield about
31,000 $\Lambda$p pairs in the foreground distribution, with
$q \le 100$~MeV/c, after applying the track-merging cut for the 6 AGeV data.
We have corrected the $\Lambda$p correlation function for: (a) the
combinatoric background ($\simeq 20$\%) included in the $\Lambda$ sample,
(b) feed-down due to the electromagnetic decay of the $\Sigma^0$ (estimated to
be $\simeq 25$\% from RQMD calculations) and (c) smearing due to momentum
resolution of the detector.

We corrected the sources for momentum resolution using two independent methods,
both leading to consistent results. In the first method, we left the
data uncorrected and modified the kernel used in the imaging analysis
to include the smearing effect, assuming an average
$\delta p$ for each $\left<p_T\right>$ and $\left<y\right>$.  This technique
will be detailed elsewhere~\cite{smearker}.
In the second method, we first corrected the correlation functions via an
iterative procedure and imaged with an unsmeared kernel.  We explain this
procedure in the next paragraph.
We determined the momentum resolution for $\pi^-$ and p from GEANT
simulations giving an average value of 2.0\% and 3.1\% for each component of
the $\pi^-$ and proton momentum resolution, respectively.  For the
$\Lambda$ momentum resolution, we extracted an average value of 4.0\%
for each component of the resolution using the width of the $\Xi^-$
mass peak ($\Xi^-$ decays to $\Lambda\pi^-$) and the $\pi^-$ momentum
resolution.

Our iterative procedure starts with model calculations (Gaussian sources)
for the particle momenta and their (unsmeared) correlations, $C_u$(q).
We then use a Monte Carlo method to smear their momenta and
use these to calculate the smeared correlation $C_s$(q).
The ratio $C_s$(q)/$C_u$(q) is used to make a first-try correction to the raw
observed data correlation C(q). This gives a first iteration unsmeared
correlation $C'_u$(q). The latter is then smeared in the
second iteration (using pairs of particle from mixed data events) to give a
second smeared correlation, $C'_s$(q). Typically, the comparison between the
raw observed C(q) in the data and the second smeared correlation function led
to a reduced $\chi^2 \simeq$ 1. The associated function $C'_u$(q) was then
taken as the unsmeared correlation for the data.
Figure~\ref{smr_unsmr} shows both smeared and unsmeared correlation functions.
The results presented in this paper are from the iterative method.

p$\Lambda$ correlations can lead to residual
correlations between primary protons and daughter protons from $\Lambda$ decays.
Wang has calculated the magnitude of this residual effect on the pp correlation
\cite{wang99}. We have determined its effect
on our pp correlations to be negligible.
This is due to (1) the intrinsically small residual
 pp correlation (maximum 3\%) arising from our observed p$\Lambda$ correlation
and (2) the small fraction of secondary protons resulting from $\Lambda$
decay ($\approx$ 6\% of the total number of protons).
The same reasoning applies for the expected perturbation of p$\Lambda$
correlations from $\Lambda\Lambda$ correlations \cite{na49} and of
$\pi^-\pi^-$ correlations from $\pi^-\Lambda$ or $\pi^-K^0$ correlations
\cite{chu02}.

\begin{figure}
\includegraphics[width=0.45\textwidth]{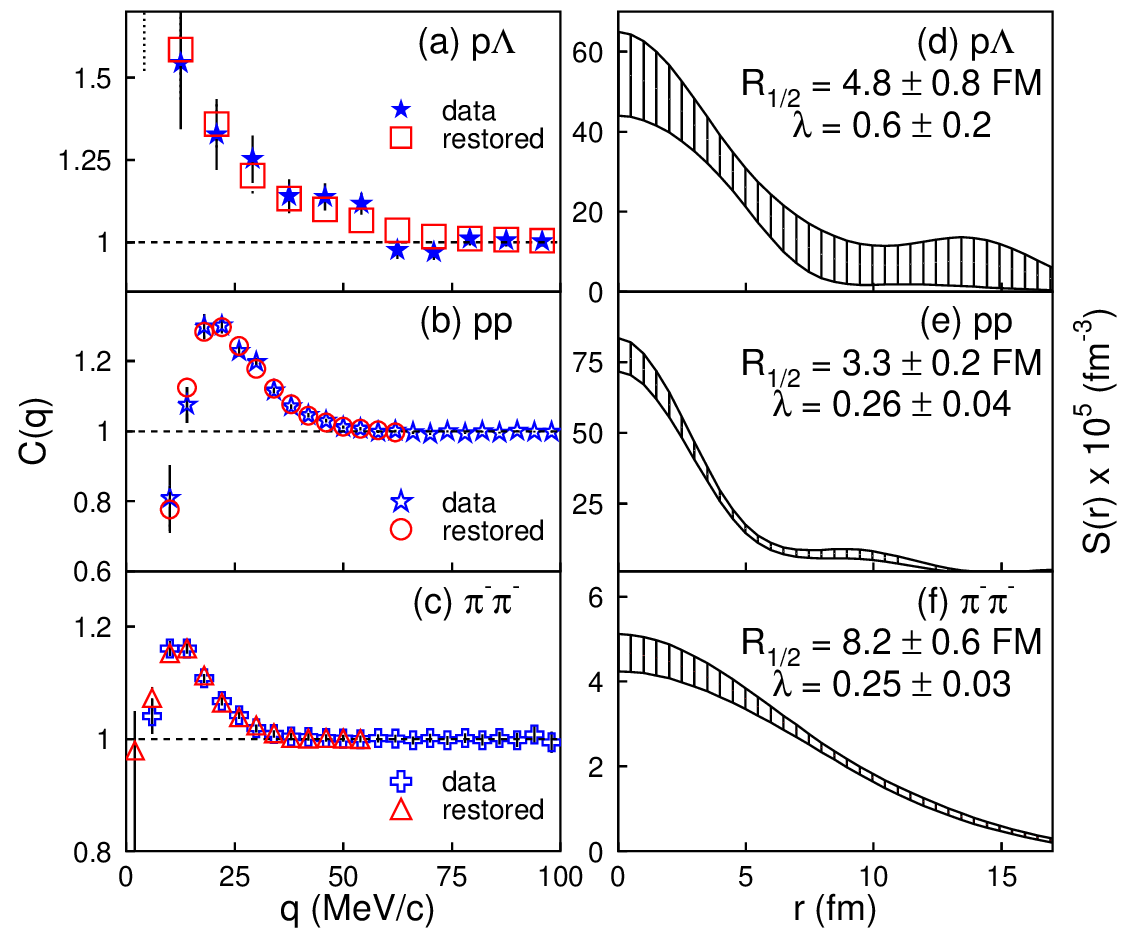}
\caption{
Correlation functions, C(q), for p$\Lambda$, pp and $\pi^-\pi^-$ pairs
are shown in panels (a), (b) and (c) respectively.
The corresponding short-range source functions, S(r), are indicated
in panels (d), (e) and (f). As a consistency test, a simulated correlation function
(open squares, circles and triangles) is recalculated from S(r). Hatched bands
show the zone of one standard deviation.
}
\label{lam_p_cor}
\end{figure}

Figures.~\ref{lam_p_cor}(a),(b),(c) show correlation functions for
$\Lambda$p, pp and $\pi^-\pi^-$ pairs respectively.
We have not corrected the pp and $\pi^-\pi^-$ correlation
functions for the Coulomb interaction
since this effect is included in the imaging procedure.
The two particle correlation and the source function are related
through the Koonin-Pratt equation~\cite{koopratt}:
\begin{equation}
  C(q)-1=4\pi\int dr r^2 K(q,r) S(r).
  \label{kpeqn}
\end{equation}
This is a linear integral equation that we may invert to obtain the
source function $S(r)$ using the techniques in Ref.~\cite{bro01}.
Here, the imaged source function $S(r)$ gives the probability of emitting a
pair of particles a distance $r$ apart in the pair c.m. frame.
The derived source functions are shown in Figs.~\ref{lam_p_cor} (d),(e),(f).
As a consistency check, we have recalculated the correlation functions from the
derived source functions and these are shown in Figs.~\ref{lam_p_cor}(a),(b),
(c) as restored correlations.

In Eq.~(\ref{kpeqn}), the kernel $K(q,r)$ encodes the FSI and is given in
terms of the final state wavefunction as $K(q,r)=\frac{1}{2}\int
d(\cos(\theta_{\bf q, r})) (|\Phi_{\bf q}({\bf r})|^2-1)$.
In this work, we used the Coulomb force for the pion and proton pairs.
Additionally, we used the Reid93 nucleon-nucleon force for
protons~\cite{reid93} and the phenomenological potential
of Bodmer and Usmani for $\Lambda$p pairs~\cite{bodusmani}.
Since the $\Lambda$p potential is not very well known,
we reanalyzed the p$\Lambda$ correlation using
a simplified kernel that depends only on the p$\Lambda$ effective
range and scattering length.  We found
no significant change in the imaged $\Lambda$p source.

The sources in Figs.~\ref{lam_p_cor} (d),(e),(f) appear Gaussian, although we
cannot definitively conclude this given the size of the error bands on the
imaged sources.  In principle, the source function can be composed of an
admixture of short and large-range emission sources \cite{bro01}.  In practice,
the shape of the correlation is strongly dominated by the short-range source
and the less-correlated pairs from any large-range source essentially dilute
the strength of the observed correlation.  Let us then assume a Gaussian
for the short-distance part of the time integrated emission function for
particle type $i$:
$D_i \sim f_i\exp{\left(-r^2/2R_i^2\right)}$.  The fraction of particles
emitted from this source is $0\le f_i\le 1$.
This choice gives us a Gaussian two-particle source function for particles
$i$ and $j$:
\begin{equation}
  S_{ij}(r) \propto \lambda_{ij}\exp{\left(-r^2/2(R_i^2+R_j^2)\right)}.
  \label{source}
\end{equation}
To ensure that we use the same source parameterization for like and unlike
pairs, we choose $R_{ij}^2=\frac{1}{2}(R_i^2+R_j^2)$.  With this choice the
emission function radius is exactly the two particle source radius if $i=j$
(i.e. $R_{ii}=R_{i}$) \cite{radii}.
Here the fraction of pairs $\lambda_{ij}$ is
related to the fraction of particles in each emission function through
$\lambda_{ij}=f_if_j$.  In all subsequent discussion, we take the
$R_{1/2}$ value (the radius at half maximum density) directly from the
sources, but we convert the source height into an equivalent Gaussian
$\lambda$.  Values for $\lambda$ and $R_{1/2}$
of the short-range sources are indicated in Fig.~\ref{lam_p_cor} (d),(e),(f).

Scanning Fig.~\ref{lam_p_cor}(d),(e),(f),
the pion source is clearly the broadest.  However, our naive expectation that
the $\Lambda p$ source would be the smallest source appears wrong.  In an effort
to understand why, we investigate the source sizes in more detail.

The results in Fig.~\ref{lam_p_cor}(c),(f) for $\pi^-\pi^-$
show that $R_{1/2}$ = 8.2$\pm$0.6 fm. This radius,
corresponding to a centrality range of b~$<$ 7 fm, is identical
to that reported by Panitkin et al for b~$<$ 5 fm\cite{pan01,rhalf}.
The $\lambda$ value of 0.25$\pm$0.03 indicates that half of the pions arise from
a source with $R_{1/2} \sim 8.2$~fm and the other half from a considerably
larger source, possibly caused by resonance decays.  
For $p\Lambda$ pairs [cf. Figs.~\ref{lam_p_cor}(a),(d)] the correlations are 
dominated by a source of smaller size ($R_{1/2}$ = 4.8$\pm$0.8 fm) comprising 
60\% $\pm$ 20\% of the $p\Lambda$ pairs.

The contrast in size with the $pp$ source is pronounced
(cf. Figs.~\ref{lam_p_cor}(b),(e)).  The pp correlations are dominated by a
very compact source $R_{1/2}$ = 3.3$\pm$0.2 fm and this source comprises only
26{\%} of the pairs or 51{\%} of the total protons.
Both in 75 AMeV $^{14}$N+$^{197}$Au \cite{verde01} 
and 200 AGeV S+Pb \cite{bro01} investigators have also 
found relatively small source sizes ($R_{1/2}= 2.9$ fm and $R_{1/2}= 3.2$ fm, 
respectively) for roughly the same fraction of protons.
From these observations, we suspect that the there is a
common feature in nucleus-nucleus collisions that span a broad range
because (1) strong collective motion focuses the source to much smaller
radii~\cite{pan00}, and (2) pp correlations are insensitive to the long distance
parts of the source, as discussed by Wang and Pratt~\cite{wan99}.
While we cannot separate these two effects in this study (or even rule out 
other more exotic causes) we comment that pions are much less focused by 
collective effects because of their much lower mass~\cite{flow}.  
Consequently, we expect the pion correlation function to probe a larger portion 
of the emission region and hence be associated with a larger source.

\begin{figure}
\includegraphics[width=0.45\textwidth]{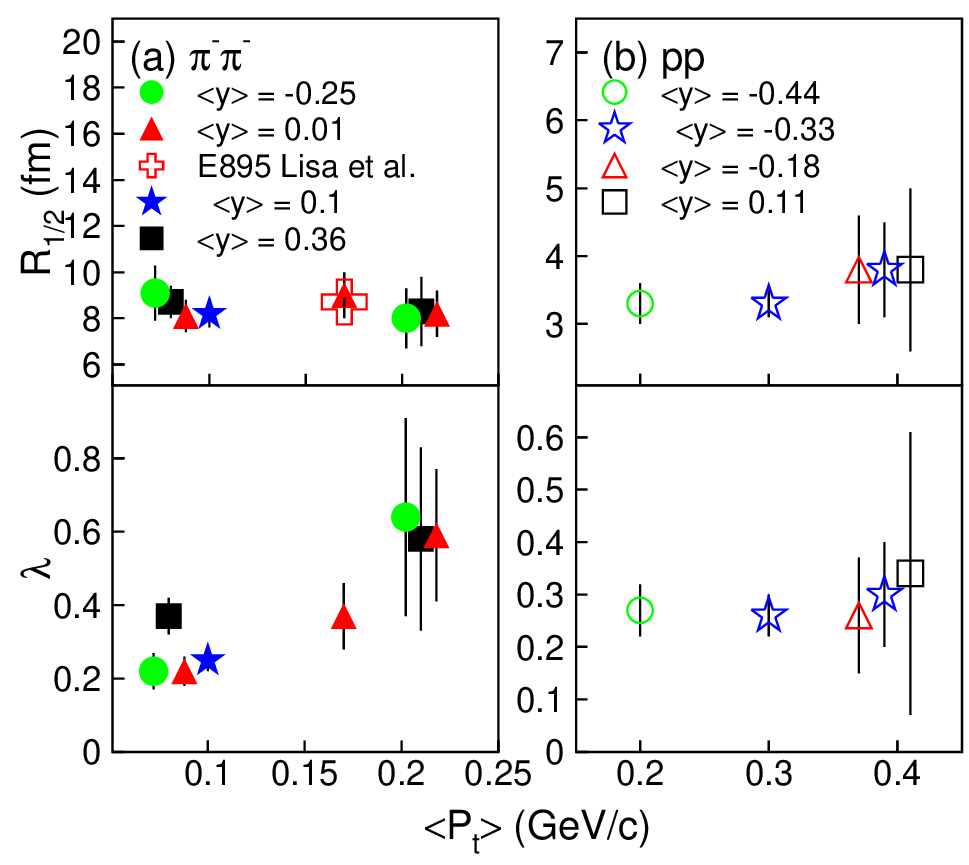}
\caption{Variation of source half-radius $R_{1/2}$ (top panels) and $\lambda$
parameter (bottom panels) for different phase space regions
identified by their mean transverse momentum $\left<P_t\right>$ and mean
rapidity $\left<y\right>$
values (shown in different symbols) for sources from (a) $\pi^-\pi^-$ and (b)
pp correlations. The open cross in figure (a) shows the $R_{1/2}$ value,
calculated from the HBT radii from Ref [7].
}
\label{rhalf_lambda_pt}
\end{figure}

Previous works \cite{lis00,ahle02} have indicated that the 3-dimensional pion radii exhibit 
$m_T$ scaling attributable to collective flow. 
$R_{1/2}$ in the pair c.m. frame can be computed from the 3-dimensional source 
radii $R_{\text{\scriptsize out}}$, $R_{\text{\scriptsize side}}$ and 
$R_{\text{\scriptsize long}}$ in the Longitudinally Comoving Source 
(LCMS) frame via
\begin{equation}
  R_{1/2}\approx 1.66\sqrt[3]{\gamma R_{\text{\scriptsize out}} 
  R_{\text{\scriptsize side}} R_{\text{\scriptsize long}}},
  \label{eqn:1d3dradii}
\end{equation}
where the Lorentz factor is $\gamma=m_T/m$.  
In the various models of $m_T$ scaling of correlation radii, the 
radii in the LCMS frame all have the same form: $R = R^0 \sqrt{T/m_T}$.
Here $T$ is the source temperature and the $R^0$'s for each radius depend 
on the details of the model in question.  Inserting this dependence into 
Eq. (\ref{eqn:1d3dradii}) gives 
\begin{equation}
  R_{1/2}\approx R^0_{1/2}\sqrt{T/m_T^{\text{\scriptsize eff}}},
  \label{eqn:1dmTscaling}
\end{equation}
where $m_T^{\text{\scriptsize eff}}=\sqrt[3]{m_T m^2}$.  This implies that 
$R_{1/2}$ for a given system scales weakly as $m_T^{-1/6}$ regardless of 
the $m_T$ scaling model as is evident in Fig. 4. 

If one assumes that $R^0_{1/2}$ and T are the same for both $\pi^-$ and proton sources
then Eq.~(\ref{eqn:1dmTscaling})predicts the ratio $R_{pp}/R_{\pi^-\pi^-}=\sqrt{m^{\text{\scriptsize eff}}_{T
\pi^-}/m^{\text{\scriptsize eff}}_{T p}}=0.4$ which is consistent with the value 
of the experimental ratio ($0.40\pm 0.04$). 

	In contrast to the $pp$ and $\pi^-\pi^-$ sources, the $p\Lambda$
source size does not show this $m_T$ scaling. Naively applying 
Eq.~(\ref{eqn:1dmTscaling}) and keeping a constant temperature and geometrical source size, 
one obtains a predicted $p\Lambda$ source size of $3.1\pm 0.2$ fm which is significantly 
smaller than the experimental value. This value cannot be accounted for via the proton-$\Lambda$ 
mass difference of $\approx$20\%. It could reflect differences between the emission 
time and the extent of collective focusing for $\Lambda$'s and protons.
In fact, the magnitude of collective flow for $\Lambda$'s is known to be
smaller than that for protons \cite{chung2001}.

As a possibility, let us assume that the underlying proton emission function is
common to both the $pp$ and $p\Lambda$ correlations.  For this assumption to 
hold, we must assume that any flow induced
focusing affects the $pp$ and $p\Lambda$ sources similarly and that we may
concentrate on only the short-range proton and $\Lambda$ sources.
In this context, we can use Eq.~(\ref{source}) and the $pp$ source
function values to extract information about the $\Lambda$ emission function.
We find that 
$f_{\Lambda}=\lambda_{\Lambda p}/\sqrt{\lambda_{pp}}=1.18\pm 0.40$ and
$R_{\Lambda}=\sqrt{2R^2_{\Lambda p}-R^2_{pp}}=5.93\pm 1.30$~fm.
Given that $f_{\Lambda}\simeq 1$, we may argue that all
$\Lambda$'s are made from this source.  
Such a moderate sized $\Lambda$ source could indicate the geometrical size of
the hot central region of the collision zone, where the energy density is high
enough for the production of the strange quarks that form the observed 
$\Lambda$'s.

In summary, we have measured small-angle correlations for $p\Lambda$, $pp$ and
$\pi^-\pi^-$ pairs produced in 6 AGeV Au+Au reactions and have analyzed them by 
the source imaging technique \cite{bro01}.  The strong differences in effective 
source radii reflect very different dynamical histories for each pair.
The $pp$ and $\pi^-\pi^-$ pairs may reflect a small homogeneity length 
caused by flow focusing the source while the $\Lambda$ emission functions 
reflect a spread out participant zone.

This work was performed under the auspices of the U.S. Department of
Energy by University of California, Lawrence Livermore National
Laboratory under Contract W-7405-Eng-48. This work was also supported by NSF
grant PHY-00-70818, U.S. DOE grant DE-FG02-88ER40412 and other grants
acknowledged in Ref.\cite{chu00b}.


\begin{thebibliography}{70}
\bibitem{radii} We use this convention so that we can directly compare the
Gaussian radii to standard pion radii.  This convention is the same as that
in Ref~\cite{pan01}.
\bibitem{rhalf} $R_{1/2} = 1.66 \times R_G$ in Ref.\cite{pan01}. Assigned
uncertainties include the absolute error in data, corrections
and methodology.  Hence, they are larger than those in Ref.\cite{pan01}.
\bibitem{flow} The ratio of thermal velocity to expansion velocity for pions
will exceed that for protons.
\end{thebibliography}
\end{document}